\documentclass[aps,pre,twocolumn]{revtex4-1}
\usepackage{graphicx}
\usepackage{dcolumn}
\usepackage{bm}
\usepackage{amsmath}
\usepackage{color}
\begin{document}

\title{Structural instability of large-scale functional networks}

\author{Shogo Mizutaka}
\email{mizutaka@ism.ac.jp}
\affiliation{School of Statistical Thinking, The Institute of
Statistical Mathematics, Tachikawa 190-8562, Japan}
\author{Kousuke Yakubo}
\email{yakubo@eng.hokudai.ac.jp} \affiliation{Department of
Applied Physics, Hokkaido University, Sapporo 060-8628, Japan}

\date{\today}
\begin{abstract}
We study how large functional networks can grow
stably under possible cascading overload failures and evaluated
the maximum stable network size above which even a small-scale
failure would cause a fatal breakdown of the network. Employing
a model of cascading failures induced by temporally fluctuating
loads, the maximum stable size $n_{\text{max}}$ has been
calculated as a function of the load reduction parameter $r$
that characterizes how quickly the total load is reduced during
the cascade. If we reduce the total load sufficiently fast
($r\ge r_{\text{c}}$), the network can grow infinitely.
Otherwise, $n_{\text{max}}$ is finite and increases with $r$.
For a fixed $r\,(<r_{\text{c}})$, $n_{\text{max}}$ for a
scale-free network is larger than that for an exponential
network with the same average degree. We also discuss how one
detects and avoids the crisis of a fatal breakdown of the
network from the relation between the sizes of the initial
network and the largest component after an ordinarily occurring
cascading failure.
\end{abstract}
\maketitle

\section{Introduction}
Numerous complex systems in nature and society can be
simplified and abstracted by describing them as networks, in
which nodes and edges represent constituent elements and their
interactions, respectively. Extensive studies
\cite{Albert02,Boccaletti06,Dorogov08} have revealed common
statistical features of real-world complex networks, such as
the small-world property \cite{Watts98}, the scale-free
property \cite{Barabasi99}, community structures
\cite{Girvan02}, and degree-degree correlations
\cite{Newman02}. In order to clarify the origin of these
features and/or properties of various dynamics on such
networks, so many network models have been proposed so far. In
most of previous network models, the number of nodes $N$,
namely the network size, is treated as an a priori given
parameter. The network size can then take any value, and, as is
often the case, the limit of infinite $N$ is taken in order to
simplify the analysis. Thus, these models implicitly assumes
that networks are stably present no matter how large networks
grow. This assumption is, however, not always valid in
real-world systems. In an ecological network representing a
closed ecosystem, for example, too many species destabilize the
ecosystem and the number of species (nodes in the ecological
network) cannot increase unboundedly \cite{Gardner70,May72}.
Also in a trading network, too many firms make the network
fragile because of unstable cartels \cite{Porter83}, increase
of financial complexity \cite{Haldane11,Heiberger14}, possible
large-scale chain-bankruptcy, and other risks \cite{Meade10}.
As in these examples, due to intrinsic instability of
large-scale networks, some sort of networks have their own
limit in sizes only below which they can be stable
\cite{Shimada14,Watanabe15}.

It is important to study limit sizes of connected networks and
find a way to control them. Such information for
\textit{functional networks} is particularly crucial, because
unstable functional networks are directly connected to the
instability of our modern society supported by them. Functions
provided by functional networks are guaranteed by global
connectivity and normal operation of each node (or edge).
However, the larger a network grows, the lower the probability
that all the nodes operate normally becomes. When failures are
caused by \textit{overloads}, even a few failures can spread to
the entire network through a cascading process, which leads the
fatal breakdown of the network \cite{Motter02,Holme02b,
Crucitti04,Buldyrev10,Gao11,Zhou12}. It is then significant to
investigate how large functional networks can grow, while
maintaining its global connectivity, by overcoming cascading
overload failures. In this paper, we evaluate the upper limit
of connected network size above which the network becomes
unstable by cascading overload failures and discuss how the
maximum stable size can be controlled. We examine uncorrelated
random networks with Poisson and power-law degree distributions
by employing the model of cascading failures triggered by
temporally fluctuating loads \cite{Mizutaka15}.

The rest of this paper is organized as follows. In
Sec.~\ref{sec:model_method}, we briefly review the model of
cascading overload failures proposed by Ref.~\cite{Mizutaka15}
and explain how we analyze the stability of networks against
cascading overload failures. Our analytical results and their
numerical confirmations are presented in
Sec.~\ref{sec:results}. Section \ref{sec:sammary} is devoted to
the summary and some remarks.

\section{Model and Methodology}
\label{sec:model_method}

\subsection{Model}
\label{subsec:model} In this section, we outline the model of
cascading failures induced by temporally fluctuating loads
\cite{Mizutaka15}. In functional networks such as power grids,
the Internet, or trading networks, some sort of ``flow"
(electric current in a power grid, packet flow in the Internet,
and money flow in a trading network) realizes their functions.
And flow, at the same time, plays a role of ``loads" in these
networks. The load on a node usually fluctuates temporally and
the node fails if the instantaneous value of the load exceeds
the node capacity. Since flux fluctuations at a node exhibit
the same scaling behavior with fluctuations of the number of
non-interacting random walkers on the node
\cite{Menezes04,Meloni08}, Kishore \textit{et al.} modeled
fluctuating loads by random walkers moving on a network and
calculated the overload probability that the number of walkers
exceeds the range allowed for a node \cite{Kishore11}. The
model of cascading failures employed in the present work
utilizes this overload probability.

In a connected and undirected network with $M_{0}$ edges, the
probability $h_{k}(w)$ that $w$ random walkers (loads) exist on
a node of degree $k$ is presented by
\begin{equation}
h_{k}(w)=\binom{W_{0}}{w}p_{k}^{w}(1-p_{k})^{W_{0}-w},
\label{eq:binom}
\end{equation}
where $W_{0}$ is the total number of walkers and
$p_{k}=k/2M_{0}$ is the stationary probability to find a random
walker on a node of degree $k$ \cite{Noh04}. This leads a
natural definition of the capacity $q_{k}$ as
\begin{equation}
q_{k}={\langle w\rangle}_{k} + m\sigma_{k},
\label{eq:capacity}
\end{equation}
where $m$ is a real positive parameter characterizing the node
tolerance, and ${\langle w\rangle}_{k}$ and $\sigma_{k}$ are
the average and the standard deviation of $h_{k}(w)$, which are
given by ${\langle w\rangle}_{k}=W_{0}p_{k}$ and
$\sigma_{k}=\sqrt{W_{0} p_{k}(1-p_{k})}$, respectively. Since
the overload probability $F_{W_{0}}(k)$ is the probability of
$w$ to exceed $q_{k}$, we have \cite{Kishore11}
\begin{eqnarray}
F_{W_{0}}(k) &=&  \sum_{w=\lfloor q_{k}\rfloor +1}^{W_{0}}
                \binom{W_{0}}{w}p_{k}^{w}(1-p_{k})^{W_{0}-w} \nonumber \\
             &=&I_{k/2M_{0}}(\lfloor q_{k}\rfloor +1,W_{0}-\lfloor q_{k}\rfloor ),
\label{eq:FWk}
\end{eqnarray}
where $I_{p}(a,b)$ is the regularized incomplete beta function
\cite{Abramowitz64} and $\lfloor x\rfloor$ denotes the largest
integer not greater than $x$.

Using the above overload probability, the cascade process of
overload failures is defined as follows \cite{Mizutaka15}:

(i) Prepare an initial connected, uncorrelated, and undirected
network $\mathcal{G}_{0}$ with $N$ nodes and $M_{0}$ edges, in
which totally $W_{0}$ random walkers exist, and determine the
capacity $q_{k}$ of each node according to
Eq.~(\ref{eq:capacity}). $W_{0}$ is set as $W_{0}=aM_{0}$,
where the parameter $a$ is the load carried by a single edge.

(ii) At each cascade step $\tau$, reassign $W_{\tau}$ walkers
to the network $\mathcal{G}_{\tau}$ at step $\tau$, where the
total load $W_{\tau}$ is given by
\begin{equation}
W_{\tau}=\left( \frac{M_{\tau}}{M_{0}} \right)^{r} W_{0}\ .
\label{eq:Wtau}
\end{equation}
Here, $M_{\tau}$ is the total number of edges in the network
$\mathcal{G}_{\tau}$ and $r$ is a real positive parameter.

(iii) For every node in $\mathcal{G}_{\tau}$, calculate the
overload probability given by
\begin{equation}
F_{W_\tau}(k_{0},k)=I_{k/2M_{\tau}}\left(\lfloor q_{k_{0}}(W_{0})\rfloor
+1, W_{\tau}-\lfloor q_{k_{0}}(W_{0})\rfloor \right),
\label{eq:Ftau}
\end{equation}
where $k_{0}$ and $k$ are the initial degree and the degree of
the node at cascade step $\tau$, and remove nodes from
$\mathcal{G}_{\tau}$ with this probability.

(iv) Repeat (ii) and (iii) until no node is removed in the
procedure (iii).

The reduction of the total load in the procedure (ii)
corresponds to realistic situations in which the total load is
reduced to some extent during a cascade process to prevent the
fatal breakdown of the network function. When a company goes
bankrupt on a trading network, for example, a large-scale chain
bankruptcy would be prevented by the reduction of the total
debt (loads) realized by financial bailout measures. The
exponent $r$ characterizes how quickly the total load decreases
with decreasing the network size, which is called the load
reduction parameter.

During the cascade, the network $\mathcal{G}_{\tau}$ might be
disconnected even though the initial network $\mathcal{G}_{0}$
is connected. In such a case, a walker on a connected component
cannot jump to other components. Therefore, the amount of
walkers on each component is conserved in the random walk
process. The overload probability then becomes dependent on how
the total load is distributed to disconnected components. Thus
the overload probability deviates from Eq.~(\ref{eq:Ftau}).
This deviation is, however, small and the effect of
disconnected components can be approximately neglected as
argued in details in Ref.~\cite{Mizutaka15}. The validity of
this approximation will be confirmed in the next section by
numerical simulations in which walkers distributed
proportionally to the number of edges in each component cannot
move to other components.

\subsection{Size of the largest component}
\label{subsec:largest} We examine the stability of a network
under cascading overload failures described above by analyzing
the size $n_{\text{f}}$ of the largest connected component in
the network after completed the cascading process. If
$n_{\text{f}}$ is very small, the initial network is considered
to be unstable. The quantity $n_{\text{f}}$ obviously depends
on the initial network size $N$, and the maximum value
$n_{\text{max}}$ of $n_{\text{f}}$ with respect to $N$ provides
the upper limit of the size of stable connected networks in a
given cascading condition. The surviving component of size
$n_{\text{max}}$ may experience further cascading failures
after a long time, but simultaneously the component can grow
during this period. In the competition between the growth and
decay processes, the component smaller than $n_{\text{max}}$
can, in substance, stably grow up to $n_{\text{max}}$.
Therefore, the connected network size fluctuates around this
maximum size $n_{\text{max}}$.

In order to calculate  the maximum stable size $n_{\text{f}}$,
we construct a master equation for the probability
$\Pi_{\tau}(k_{0},k)$ that a randomly chosen node has the
degree $k$ at cascade step $\tau$ and the initial degree $k_0$.
It is convenient to introduce another probability
$\phi_{\tau}(k)$ of a node adjacent to a randomly chosen node
of degree $k$ to experience an overload failure at cascade step
$\tau$. This probability is independent of $k$ for uncorrelated
networks and is given by \cite{Mizutaka15}
\begin{equation}
\phi_{\tau} =\sum_{k_{0}}\sum_{k'=1}^{k_{0}}\frac{k'\Pi_{\tau}(k_{0},k')}{\langle k\rangle_{\tau}}F_{W_{\tau}}(k_{0},k'),
\label{eq:phi2}
\end{equation}
where $\langle k\rangle_{\tau}$ is the average degree of
$\mathcal{G}_{\tau}$. We then formulate the master equation for
$\Pi_{\tau}(k_{0},k)$ as
\begin{widetext}
\begin{equation}
\Pi_{\tau}(k_{0},k)=\sum_{k'\ge k}\Pi_{\tau-1}(k_{0},k')
\Biggl\{\binom{k'}{k}\phi_{\tau-1}^{k'-k}(1-\phi_{\tau-1})^{k}
[1-F_{W_{\tau-1}}(k_{0},k')]+\delta_{k0}F_{W_{\tau-1}}(k_{0},k')\Biggr\}.
\label{eq:Pikk}
\end{equation}
\end{widetext}
In this equation, we do not remove overloaded nodes actually
but leave them in the system as zero-degree nodes, which makes
the theoretical treatment easier. The right-hand side of this
equation represents the probability that a degree-$k'$ node in
$\mathcal{G}_{\tau-1}$ becomes a node of degree $k$ at cascade
step $\tau$. The first term describes the situation that the
degree-$k'$ node does not experience an overload failure and
$k'-k$ nodes adjacent to this node fail. The second term stands
for the case that the degree-$k'$ node itself fails and becomes
a zero-degree node. Solving numerically Eq.~(\ref{eq:Pikk})
with the aid of Eq.~(\ref{eq:phi2}), we can calculate
$\Pi_{\tau}(k_{0},k)$ iteratively starting from
$\Pi_{0}(k_{0},k)=P_{0}(k)\delta_{kk_{0}}$, where $P_{0}(k)$ is
the degree distribution function of $\mathcal{G}_{0}$.
According to the procedure (iv), we stop this iterative
calculation at step $\tilde{\tau}$ satisfying the condition
\begin{equation}
\sum_{k,k_{0}}F_{W_{\tilde{\tau}}}(k_{0},k)\Pi_{\tilde{\tau}}(k_{0},k)<\frac{1}{N},
\label{eq:eqcond}
\end{equation}
which implies that the expectation number of overloaded nodes
becomes less than unity.

We can obtain the largest connected component size
$n_{\text{f}}$ at the final cascade step $\tilde{\tau}$ from
the degree distribution $P_{\tilde{\tau}}(k)$ of
$\mathcal{G}_{\tilde{\tau}}$ which is given by
$P_{\tilde{\tau}}(k)=\sum_{k_{0}\ge
k}\Pi_{\tilde{\tau}}(k_{0},k)$. Employing the generating
function formalism, $n_{\text{f}}$ is calculated by
\cite{Newman01}
\begin{equation}
n_{\text{f}}=N \left[1-\sum_{k}P_{\tilde{\tau}}(k)u^{k}\right],
\label{eq:ntau}
\end{equation}
where $u$ is the smallest non-negative solution of
\begin{equation}
u=G_{1}(u),
\label{eq:u}
\end{equation}
and $G_{1}(x)$ is the generating function of the remaining
degree distribution, which is defined by
\begin{equation}
G_{1}(x)=\frac{1}{\langle k\rangle_{\tau}}\sum_{k}(k+1)P_{\tilde{\tau}}(k+1)x^{k}.
\label{eq:G1}
\end{equation}
It should be noted that Eq.~(\ref{eq:ntau}) does not mean that
$n_{\text{f}}$ is proportional to $N$ because
$1-\sum_{k}P_{\tilde{\tau}}(k)u^{k}$ depends on $N$.

\section{Results}
\label{sec:results} 
\begin{figure}[ttt]
\begin{center}
\includegraphics[width=0.48\textwidth]{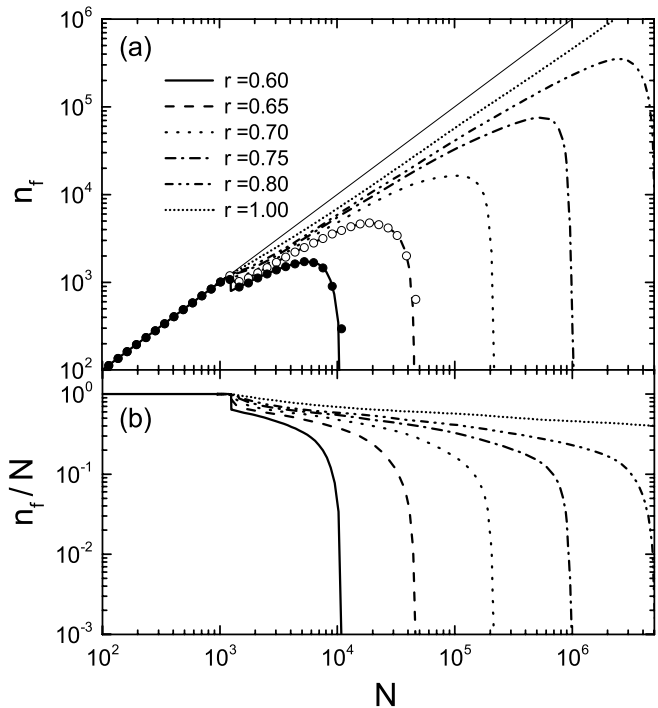}
\caption{(a) Largest component size $n_{\text{f}}$ after
completing cascading overload failures as a function of the
size $N$ of initial ERRGs, for various values of
$r$. The thick lines indicate $n_{\text{f}}$ calculated by
Eq.~(\ref{eq:ntau}) and filled and open circles
on thick lines show the results
obtained by the numerical simulation described in
the main text. The thin straight line is a guide
to the eyes for $n_{\text{f}}=N$. (b) Relative largest
component size $n_{\text{f}}/N$ as a function of $N$. Lines
have the same meanings as those in (a). All the
results are calculated for ${\langle k\rangle}_{0}=5.0$,
$m=4.0$, and $a=2.0$.}
\label{fig1}
\end{center}
\end{figure}

First, we calculated $n_{\text{f}}$ for the
Erd\H{o}s-R\'enyi random graph (ERRG) as an initial network
$\mathcal{G}_{0}$. In this case, the binomial degree
distribution function for $\mathcal{G}_{0}$ is given by
\begin{equation}
P_{0}(k)=\binom{N-1}{k}p^{k}(1-p)^{N-k-1},
\label{binomial}
\end{equation}
where $p={\langle k\rangle}_{0}/N$. In this work, we fix the
initial average degree as $\langle k\rangle_{0}=5.0$. Although
initial networks having this average degree are not completely
connected with isolated nodes at a very low rate, this does not
affect our conclusion. Figure \ref{fig1} shows $n_{\text{f}}$
as a function of the initial network size $N$ for various
values of the load reduction parameter $r$. For these results,
the node tolerance parameter $m$ and the load carried by a
single edge, $a$, are chosen as $m=4.0$ and $a=2.0$. The lines
in Fig.~\ref{fig1} represent $n_{\text{f}}$ calculated by
Eq.~(\ref{eq:ntau}) and the symbols indicate the results
obtained by numerical simulations performing faithfully the
cascade process from (i) to (iv) described in
Sec.~\ref{subsec:model}. In the numerical simulation, the
overload probability at cascade step $\tau$ is calculated under
the condition that random walkers cannot jump to other
components. Namely, instead of Eq.~(\ref{eq:Ftau}), we adopt
the overload probability of a node in the $\alpha$-th component
given by
\begin{equation}
F_{W_{\tau}^{\alpha}}(k_{0},k)=I_{k/2M_{\tau}^{\alpha}}
\left(\lfloor q_{k_{0}}(W_{0})\rfloor +1, W_{\tau}^{\alpha}-\lfloor q_{k_{0}}(W_{0})\rfloor \right),
\label{eq:Ftaualpha}
\end{equation}
where $M_{\tau}^{\alpha}$ is the number of edges in the
$\alpha$-th component of $\mathcal{G}_{\tau}$ and
$W_{\tau}^{\alpha}=(M_{\tau}^{\alpha}/M_{\tau}) W_{\tau}$. The
remarkable agreement between the symbols and the lines suggests
that our approximation by Eq.~(\ref{eq:Ftau}) is quite
accurate.

\begin{figure}[tttt]
\begin{center}
\includegraphics[width=0.48\textwidth]{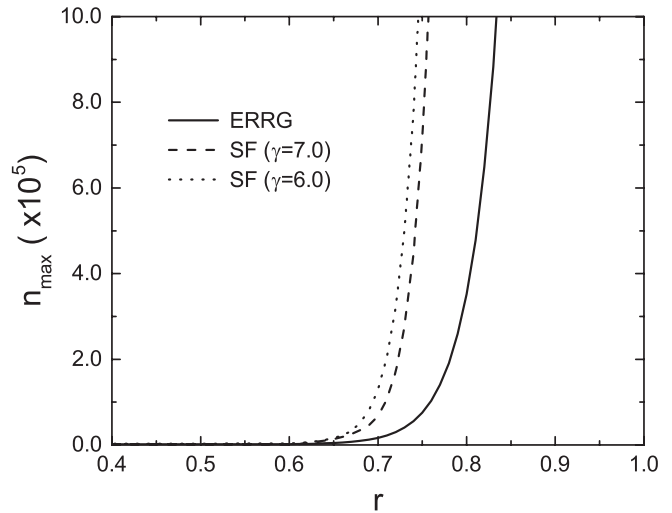}
\caption{ Maximum stable size $n_{\rm max}$ as a
function of the load reduction parameter $r$. Three lines
represent the results for the ERRG (solid line) and SF networks
with the degree distributions $P_{0}(k)$ given by
Eq.~(\ref{cauchy}) with $\gamma=7.0$ (dashed line) and $6.0$
(dotted line). The minimum degree $k_{\text{min}}$ in
Eq.~(\ref{cauchy}) is set as $1$. All the results are
calculated for ${\langle k\rangle}_{0}=5.0$, $m=4.0$, and
$a=2.0$.}
\label{fig2}
\end{center}
\end{figure}
The quantity $n_{\text{f}}$ shown in Fig.~\ref{fig1} is exactly
equal to $N$ as long as $N<N^{*}\ (\approx 10^{3})$, regardless
of the value of $r$. This implies that the network never
experiences overload failures until the network grows up to
$N^{*}$. Thus, $N^{*}$ is determined by
\begin{equation}
N^{*} = \frac{1}{\sum_{k}F_{W_{0}}(k)P_{0}(k)},
\label{eq:N*}
\end{equation}
which does not depend on $r$. When the network grows larger
than $N^{*}$, it starts to decay by initial failures and
subsequent avalanche of failures. The largest component size
$n_{\text{f}}$ after the cascade then becomes smaller than $N$,
but still increases with $N$, at least unless $N$ is much
larger than $N^{*}$. For $r\le 0.8$, when $N$ exceeds a certain
value $N_{\text{c}}(r)$, $n_{\text{f}}$ rapidly decreases with
$N$. Therefore, $n_{\text{f}}$ becomes maximum at
$N=N_{\text{c}}$. This maximum value of $n_{\text{f}}$ is
nothing but $n_{\text{max}}$ mentioned at the beginning of
Sec.~\ref{subsec:largest}. Figure \ref{fig1} clearly shows that
the maximum stable size $n_{\text{max}}$ is an increasing
function of $r$. This is because a large value of $r$, namely a
rapid decrease of the total load $W_{\tau}$ during the cascade,
prevents large-scale cascading failures in our model.

The $r$ dependence of $n_{\text{max}}$ is closely related to
the percolation transition by cascading overload failures. As
pointed out by Ref.~\cite{Mizutaka15}, there exists a critical
value $r_{\text{c}}$ above which the largest component size
diverges in proportion to $N$ in the thermodynamic limit.
Thus, $n_{\text{f}}$ goes to infinity as $N\to \infty$ for
$r\ge r_{\text{c}}$, which implies the absence (divergence) of
$n_{\text{max}}$. On the other hand, for $r<r_{\text{c}}$,
$n_{\text{f}}$ is finite and varies with $N$ to be maximized
at $N_{\text{c}}$ as mentioned above. As a consequence,
$n_{\text{max}}$ increases with $r$ for $r<r_{\text{c}}$ and
diverges at $r=r_{\text{c}}$. A finite-size scaling analysis
\cite{Stauffer91} predicts that $n_{\text{max}}$ for
$r<r_{\text{c}}$ behaves as
$n_{\text{max}}\propto |r-r_{\text{c}}|^{\beta-\nu^{*}}$ if $r$
is close enough to $r_{\text{c}}$, where the correlation volume
exponent $\nu^{*}$ and the order parameter exponent $\beta$
characterize $N_{\text{c}}$ and $n_{\text{max}}/N$ as
$N_{\text{c}}\propto |r-r_{\text{c}}|^{-\nu^{*}}$ for large
enough $N_{\text{c}}$ and $n_{\text{max}}/N\propto
(r-r_{\text{c}})^{\beta}$ for large enough $N$, respectively.
Such a behavior is demonstrated by the solid line in
Fig.~\ref{fig2} for the ERRG.  The result suggests that the
load control during a cascade is crucial to realize large
functional networks.
\begin{figure}[tttt]
\begin{center}
\includegraphics[width=0.48\textwidth]{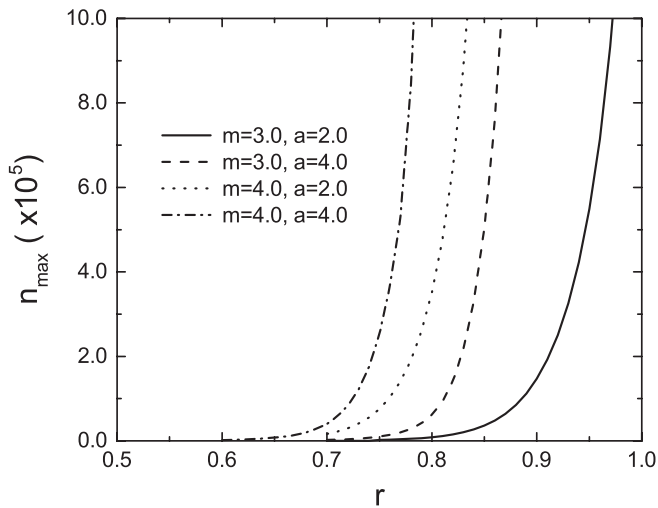}
\caption{Maximum stable size $n_{\rm max}$ for the
ERRG with $\langle k\rangle_{0}=5.0$ as a function of the load
reduction parameter $r$. Four lines represent the results for
different combinations of the node tolerance parameter $m$ and
the load carried by a single edge $a$.}
\label{fig3}
\end{center}
\end{figure}

How is the maximum size $n_{\text{max}}$ affected by properties
of the initial network $\mathcal{G}_{0}$? To clarify the
influence of the degree inhomogeneity in $\mathcal{G}_{0}$, we
first calculate $n_{\text{max}}$ for scale-free (SF) networks
with the degree distribution given by
\begin{equation}
	P_{0}(k)=
\begin{cases}
	\displaystyle 0 & \text{if }k<k_{\rm min} \\
	\displaystyle \frac{c}{k^\gamma+d^\gamma} & \text{if }k\ge k_{\rm min} ,
\end{cases}
\label{cauchy}
\end{equation}
where $d$, $\gamma$, $k_{\min}$, and the normalization constant
$c$ are real positive constants. The degree distribution has
asymptotically a power-law form, i.e., $P(k)\sim k^{-\gamma}$
for $k \gg d$. The average degree $\langle k\rangle_{0}$ can be
controlled by $d$ for a specific value of $\gamma$. The dashed
and dotted lines in Fig.~\ref{fig2} represent the results for
$\gamma=7.0$ and $\gamma=6.0$, respectively. Although these
values of $\gamma$ are not sufficiently small due to the
restriction of computational resources to calculate
$P_{\tau}(k)$ iteratively and obtain $n_{\text{f}}$ in
Eq.~(\ref{eq:ntau}), the maximum stable sizes for the SF
networks are obviously greater than that for the ERRG. This
implies that an SF network is more stable than the ERRG with
the same average degree, which is consistent with the previous
result showing the robustness of SF networks against cascading
overload failures \cite{Mizutaka15}. Next, we calculate
$n_{\text{max}}$ for several combinations of the node tolerance
parameter $m$ and the load carried by a single edge $a$. The
results depicted in Fig.~\ref{fig3} clearly indicate that
$n_{\text{max}}$ increases with both $m$ and $a$. It is obvious
that the larger the node tolerance parameter $m$, the more
stable the network consisting of tolerant nodes becomes. The
monotonous increase of $n_{\text{max}}$ with $a$ is due to the
fact that $F_{W_\tau}(k_{0},k)$ given by Eq.~(\ref{eq:Ftau}) is
a decreasing function of $a$.

It is interesting to notice that the maximum stable size
$n_{\text{max}}$ shown in Fig.~1(a) is roughly proportional to
$N_{\text{c}}$. In fact, the ratio
$n_{\text{max}}/N_{\text{c}}$ is about $0.1$ to $0.15$
independently of $r$, unless $r$ is close to $r_{\text{c}}$
\cite{comment_critical}. This ratio is also insensitive to the
parameters $m$ and $a$, as well as the degree inhomogeneity.
Furthermore, as shown by Fig.~1(b), the ratio $n_{\text{f}}/N$
slowly decreases with $N$ if $n_{\text{f}}/N \gtrsim
n_{\text{max}}/N_{\text{c}}$ but turns to a rapid decrease when
$n_{\text{f}}/N$ becomes less than
$n_{\text{max}}/N_{\text{c}}(\sim 0.1)$. These empirical facts
give us significant information about the stability of the
network. If the size of the largest component after cascading
overload failures falls close to $10\%$ to $15\%$ of the size
of the network before the cascade, the network is in immediate
danger of a fatal breakdown. In order to accomplish further
stable growth of the network, we need to raise the load
reduction parameter $r$. Of course, the value of
$n_{\text{max}}/N_{\text{c}}$ is peculiar to the present
cascade model. However, it has been found that qualitative
properties of our model are robust against changes in details
of the model as long as failures are induced by temporally
fluctuating loads \cite{Mizutaka15}. Therefore, even for a
real-world functional network, the ratio $n_{\text{f}}/N$ is
supposed to decrease drastically with $N$ when this ratio falls
below a certain value. Our results suggest that we must take
measures to prevent a fatal breakdown of a functional network
if the decreasing rate of $n_{\text{f}}/N$ with increasing the
network size becomes higher than its ordinary value.

\section{Conclusions}
\label{sec:sammary} We have studied how large a functional
network exposed to cascading overload failures can grow stably
and evaluated the maximum stable size $n_{\text{max}}$ above
which the network would face the crisis of a fatal breakdown.
To this end, we employed the model of cascading overload
failures triggered by fluctuating loads \cite{Mizutaka15} which
is described by random walkers moving on the network
\cite{Kishore11}. In this model, how quickly the total load is
reduced during the cascade to prevent the fatal breakdown is
quantified by the load reduction parameter $r$. The maximum
stable size $n_{\text{max}}$ was calculated by using the
generating function technique and solving the master equation
for the probability $\Pi_{\tau}(k_{0},k)$ that a randomly
chosen node has the degree $k$ at cascade step $\tau$ and the
initial degree $k_0$. Our results show that $n_{\text{max}}$ is
an increasing function of $r$ and diverges at a certain value
$r_{\text{c}}$. This implies that the faster the total load is
reduced during the cascade, the larger the network can grow,
and we can realize an arbitrarily large network if the total
load is sufficiently quickly reduced ($r\ge r_{\text{c}}$). It
has been also clarified that the degree inhomogeneity improves
stability of the network. More precisely, for a given
$r(<r_{\text{c}})$, $n_{\text{max}}$ for a scale-free network
is larger than that for the Erd\H{o}s-R\'enyi random graph with
the same average degree. Furthermore, from the empirical
relation between $n_{\text{max}}$ and the network size
$N_{\text{c}}$ giving $n_{\text{max}}$, we argued how one
detects and avoids the crisis of the network breakdown. The
present results suggest that a certain relative size of the
largest component size after cascading failures could be a
signature for the impending network collapse. For further
stable growth of the network, a more rapid reduction of the
total load is required during a cascade of overload failures.

In this paper, we investigated only uncorrelated networks,
while most of real-world functional networks have correlations
between nearest neighbor degrees. For a correlated network, the
probability $\phi_{\tau}$ of a node adjacent to a randomly
chosen node of degree $k$ to experience an overload failure at
cascade step $\tau$ depends on $k$. This is in contrast to the
case of uncorrelated networks, where $\phi_{\tau}$ given by
Eq.~(\ref{eq:phi2}) is independent of $k$. Thus, the analysis
becomes much more complicated for correlated networks than the
present study. However, we suppose that our conclusion does not
change qualitatively though $n_{\text{max}}$ depends on the
strength of the degree correlation. A positive (negative)
degree correlation simply makes a network robust (fragile)
against various types of failures
\cite{Goltsev08,Shiraki10,Ostilli11, Tanizawa12,Tan16}. Thus,
it seems plausible that also for our failure dynamics the
degree correlation merely shifts the value of $n_{\text{max}}$
upward or downward. Nevertheless, the relation between the
stability of a functional network and its size must be strongly
affected by the model of cascading failures. We hope that the
problem of spontaneous instability in largely grown networks
will be studied more extensively in diverse ways and models.

\begin{acknowledgements}
This work was supported by a Grant-in-Aid for Scientific
Research (No.~16K05466) from the Japan Society for the
Promotion of Science.
\end{acknowledgements}


\begin{thebibliography}{99}
\bibitem{comment_critical} If $r$ is close enough to $r_{\text{c}}$, $n_{\text{max}}/N_{\text{c}}$ varies with $N_{\text{c}}$ according to $n_{\text{max}}/N_{\text{c}}\propto N_{\text{c}}^{-\beta/\nu^{*}}$.
\end{thebibliography}
\end{document}